# Increasing Returns to Scale, Dynamics of Industrial Structure and Size Distribution of Firms[†]

*Ying Fan, Menghui Li, Zengru Di[*]*

*Department of Systems Science, School of Management, Beijing Normal University, Beijing 100875, China*

**Abstract**

A model is presented of the market dynamics to emphasis the effects of increasing returns to scale, including the description of the born and death of the adaptive producers. The evolution of market structure and its behavior with the technological shocks are discussed. Its dynamics is in good agreement with some empirical "stylized facts" of industrial evolution. Together with the diversities of demand and adaptive growth strategies of firms, the generalized model has reproduced the power-law distribution of firm size. Three factors mainly determine the competitive dynamics and the skewed size distributions of firms: 1. Self-reinforcing mechanism; 2. Adaptive firm grows strategies; 3. Demand diversities or widespread heterogeneity in the technological capabilities of different firms.

## Ⅰ. INTRODUCTION

In the past decades, the evolutionary perspective has contributed lot to the economics (Arthur, Durlauf, and Lane, 1997; Dosi, Nelson, 1994; Aruka, 2001). The economy is studied as an evolving complex system with many features in complexity such as nonlinear interactions and emergent properties. In reality, the economic system consists

[†] The paper has been presented in the 1st Sino-German workshop on the evolutionary economics, March, 2004, Beijing, China. The authors thank all the participants for their helpful comments and suggestions. The work is supported by the NSFC under the grant No. 70371072 and grant No. 79990580.

[*] Corresponding author: Tel.: +86-10-58807876, Fax: +86-10-58800141
Electronic address: zdi@bnu.edu.cn

of many adaptive agents. They learn from each other, and their values may be influenced by others' values and actions. These interactions between agents may be simple and local, but they may have important economic consequence related with the emergence of global structure. Actually the organizations in different levels usually emerge from the local processes or individual interactions. These organizations include division of labor, specialization, collective regulation, units at any given level, and industrial structure discussed in this paper. The basic approaches in the studying of complexity, including the theories of self-organization and the approach of agent-based modeling (Judson, 1994; Barr, Saraceno, 2002; Lomi, Larsen, 2001), are helpful to understand the evolution of social behavior, especially the formation of global structure in the economic and biological systems (Bonabeau, Theraulaz, *et al,* 1997; Gordon, 1997; Arthur, 1995)

Industrial structure is a very important topic in macroeconomics. Because it is closely related to the dynamics of firms and market, so it is also an interesting area for evolutionary economic research (Dosi, Nelson, 1994; Kwasnicki, Kwasnicka, 1992; Peretto, 1999; Peretto P. F., 1999a, b). From the empirical studies, we have found that there are several "stylized facts" that related to the processes of industrial evolution (see Winter, Kaniovski, Dosi, 2003; Bloch, 2004): i. Skewed firm size distributions, ii. Failure of the law of proportionate effect. One observes wide variations in the rates of growth of firms, both cross-sectionally and over time; iii. Supply shocks do bear effect on aggregate prices, which in turn influence the opportunities of survival and growth of each firm; iv. Slow adjustment of industry structure, especially after sudden endogenously generated shakeout. There are lots of works that have been done from different perspective in accordance with these empirical facts, such as innovation and competition (Kwasnicki, Kwasnicka, 1992), endogenous technological change (Peretto, 1999a, b), diffusion processes (Hashemi, 2000), stochastic industrial dynamics with heterogeneous technology (Winter, Kaniovski, Dosi, 2003), and an evolutionary approach focus on the relationship between survival of new firms and the intensity of the selection process (Reymondon, 1998). Several concepts, such as demand and supply conditions, stochastic influences, sunk costs, first-mover advantages, strategies, and methods including evolutionary theory, agent-based modeling and nonlinear dynamics are known to be helpful for analyzing industrial structure (Bloch, 2004).

In this paper, we focus on the effects of self-reinforcing mechanism on industrial structure introduced by increasing returns to scale (IRS). Increasing returns is an important reason for nonlinear interactions in economic system (Arthur, 1988). When there is IRS, the firm would have different optimal investment strategies (Wagener, 2003). And IRS has important effects on the industrial structure (Gustavasson, 2002). Actually, when industrial structure is described by firm size, number of firms and their shares in the market, IRS is a crucial factor that determines the structure. In the theory of contestable markets advanced by William J. Baumol (1983, 1988), it is revealed that industrial structure is determined by the scale effect and other factors such as cost of entry and exit. For a given production function $Y = f(\vec{X})$, where $\vec{X} = \{x_1, x_2, x_3, ..., x_n\}$ is the vector of input factors, the returns on scale $\eta$ is given by:

$$\eta = \sum_{j=1}^{n} f_j(\vec{X}) x_j / f(\vec{X}), \qquad (1)$$

where $f_j(\vec{X}) = \partial f(\vec{X}) / \partial x_j$. $\eta$ gives the change of production when all factors change with same ratio. When $\eta=1$, the enterprise has the best scale. For a given market, if we use the number of producers $N$ in the market to describe its structure, we could find that $N$ is also determined by $\eta=1$. Let's consider a simple case of production. An industry can use total factors $\vec{X} = \{x_1, x_2, x_3, ..., x_n\}$ to produce final product $Y$. In the case of variant returns to scale, we try to get maximum product by dividing the whole industry in to $N$ homogeneous factories. Then the total production of the industry is $Y = Nf(\vec{X}/N)$. So we get the following optimization problem:

$$\max_{N \geq 1} Nf(\vec{X}/N). \qquad (2)$$

Let $\partial Y / \partial N = 0$, we get:

$$f(\vec{X}/N) - \sum_{j=1}^{N} \frac{x_j}{N} \frac{\partial f}{\partial (x_j/N)} = 0. \qquad (3)$$

Compare with the formula (1), this is exactly the condition of unit return on scale. So the optimal structure of industry is directly related to the scale effect.

We discussed the industrial structure based on the above concepts. First we investigated the market dynamics with singular product. To simplify the discussion, considering a contestable market with potential demand *D* (variable with changing of price), we assume that the only factor that determines the firm's entry probability is the shortage of supply. If the current production can't match the demand, potential producer could enter the market without any entry or exit cost except a certain amount of initial fixed capital input. Every producer is a homogeneous adaptive agent. They use a production function with variant returns to scale to produce the final product. If the revenue is enough to cover the capital depreciation, they always try to reach the best scale through the capital accumulation. And if the revenue can't cover the depreciation so that the capital stock decreases and finally less that the initial capital stock, the producer will crash and exit from the market. The price of final product adjusts itself to the difference between demand and supply. According to the above ideas, a simply discrete dynamical model including the description of the born and death of the adaptive agents is constructed. Through the numerical simulation, we found that the model could give rise to the basic phenomena of nonlinear economic dynamics. There are multiple final equilibrium points. Which one would the system achieve is path dependence. The system may stabilize in an ineffective state. With the technical progress and the related variation on scale effect, the scale of each producer and their share in the market will be changed. And there is threshold effect for the technical progress. Only when the technical progress exceeds a certain critical point, could it lead to the dramatic change of the industrial structure. Some producers with lower level of technology will crash, and there will be fewer producers with larger scale in the market at last. As the results of technical progress and the competition, the price of final product will decrease and the total demand will be enlarged. All these results give a nice description for the evolution of industrial structure.

Finally, we have taken the diversity of demand into account and constructed a generalized model to describe the competitive dynamics with the evolution of the market. Rather than firms grow investment to expand productive capacity to reach the best scale, the firms now have other two strategies for the growth. One is that they can establish new plant and invest in new product. The other is that they can invest in R&D sector to improving their technology which is related directly with the IRS of production. The simulation results give the final steady distribution of the firm size. It is in good agreement with the empirical results qualitatively. From this model and the numerical simulations, we addressed three main factors that determine the competitive dynamics and the skewed distributions of firm size: 1. Self-reinforcing mechanism caused by increasing returns, that may results in first-mover advantages and path-dependent; 2. Firm strategies change adaptively as the firm grows and competition evolves, including investment in new products and technical progress; 3. Demand diversities or widespread heterogeneity in the technological capabilities of different firms.

The model is identified in detail in Section II. The market model with only one product and the corresponding results of simulation and analysis are presented. The results are mainly the evolution of the market from a given initial conditions. The limit states of $t \to \infty$ correspond to the final equilibrium states. Section III describes the generalized model with demand diversity. The final distributions of the firm size under different set of parameters are given. Some concluding remarks and suggestions for further research are given in Section IV.

## II. MARKET DYNAMICS WITH ONE PRODUCT

### A. The Model

Let's consider an economy with only one final product $Y$. The total demand $D$ could be produced by different firms. The supply is given by $Y = \sum_i Y_i$. The following are the dynamical description of the economy.

1. Demand

The demand $D$ is only related to the product price of last time period. The relationship is given by the following sigmoid function:

$$D(t) = D_0 (1 - \frac{1}{a} \frac{p(t-1)}{p(t-1)+b}). \tag{1}$$

The function gives two limits for the demand. When price equals zero, the demand reaches is maximum that is $D=D_0$. Even if the price is unreasonable large (tend to infinity), the demand keeps a finite amount: $D=D_0(1-1/a)$ ($a>1$). That means the price elasticity of the demand tends to small in the two ends of the price.

2. Supply

A). Producer: There are a number of firms in the market. At any time period, each firm produces the final product $Y_i$ with two factors: capital $K_i$ and labor $L_i$ by following production function (Puu, 1997):

$$Y_i(t) = \alpha(\beta_i(K_i(t)+L_i(t))^2 - K_i(t)^3 - L_i(t)^3). \qquad (2)$$

From the definition of return on scale ($\eta$) given in sector 1:

$$\eta = \sum_{j=1}^{n} f_j(\vec{X})x_j / f(\vec{X}),$$

where $\vec{X} = \{x_1, x_2, x_3, ..., x_n\}$, and $f_j(\vec{X}) = \partial f(\vec{X})/\partial x_j$, we can get $\eta$ for the above production function:

$$\eta_i = 2 - \alpha(K_i^3 + L_i^3)/Y_i.$$

For simplicity and without lose any generality, we assume that capital and labor are always input in proportional or more specifically in same size. So the production function becomes:

$$Y_i(t) = \alpha(4\beta_i K_i(t)^2 - 2K_i(t)^3), \qquad (3)$$

and

$$\eta_i = 2 - 2\alpha K_i^3 / Y_i.$$

For our simple case we know that when $K_i=\beta_i$, we have $\eta_i=1$, and the corresponding product is $2\alpha\beta_i^3$. So the parameter β describes the scale effect of the product.

B). Returns of the producer. Suppose that the price of labor and capital are respectively $p_l$ and $p_k$. So the cost of producer is:

$$C_i(t) = p_k K_i(t) + p_l L_i(t). \qquad (4)$$

Then for a given price of final product, the revenue of the firm is determined by:

$$R_i(t) = p(t)Y_i(t) - C_i(t).  \tag{5}$$

If the number of firms is $N(t)$, the total product, total returns, and total costs are given by:

$$Y(t) = \sum_{i=1}^{N(t)} Y_i(t), R(t) = \sum_{i=1}^{N(t)} R_i(t), C(t) = \sum_{i=1}^{N(t)} C_i(t)  \tag{6}$$

C). Growth of factors. Considering the saving of returns and the depreciation of capitals, the change of factors is:

$$K_i(t) = K_i(t-1) + s_i(t-1)R_i(t-1) - \delta K_i(t-1),  \tag{7}$$

where $\delta$ is the depreciation rate and $s_i(t-1)$ is the saving coefficient. Here we assume that the producer is an adaptive agent. He changes his saving rate from time to time in order to maintain the best scale of the firm:

$$s_i(t) = s_0(\beta - (1-\delta)K_i(t))/\beta.  \tag{8}$$

3. Price adjustment

We take the normal mechanism for the price adjustment. The price of the final product grows up when demand is over supply and it goes down when supply is over demand. This mechanism is given by the following formula:

$$p(t) = p(t-1) + \lambda \frac{D(t-1) - Y(t-1)}{D(t-1)} p(t-1), \quad \text{when } D(t\text{-}1) < Y(t\text{-}1),$$

$$p(t) = p(t-1) + \lambda \frac{D(t-1) - Y(t-1)}{D(t-1)} (p_0 - p(t-1)), \text{when } D(t\text{-}1) \geq Y(t\text{-}1).  \tag{9}$$

Here the change rate of the price is modulated by 0 when price goes down and by a maximum price $p_0$ when price goes up.

4. Born and death of the enterprises.

The market is always contestable. Either a new producer enters the market or not is only determined by the demand and supply. If the supply is already over the demand, i.e. $D(t\text{-}1) < Y(t\text{-}1)$, the born probability for new firm is 0. And if the supply can't meet the demand, some new producers should be born into the market. We assume that at the beginning of every time period, the probability for the born of new producers is proportional to the last gap between demand and supply:

$$q(t) = (D(t-1) - Y(t-1))/D(t-1), \text{ when } D(t-1) > Y(t-1). \tag{10}$$

In our numerical simulation, we assume that the number of new producer is chose randomly from 1 to 4. For each new firm, it has an initial capital $K_0$. If in the competition of the market the revenues of the firm can not cover the depreciation of the exist capital and at last leads the capital smaller than $K_0$, the firm will be dead from the market.

Under certain initial conditions, the above dynamical system can give us a nice description of the evolution of market structure.

## B. Simulation Results

In our following simulations, we set parameters as: $D_0$=6000, $a$=2.0, $b$=0.05, $\alpha$=2, $p_k$=$p_l$=1, $p_0$=0.15, $\lambda$=0.1, $\delta$=0.05, and the initial capital for new firm is $K_0$=1. The initial expected price for the product is $p(0)$=0.1. And the corresponding potential demand is $D(0)$=4000. With this initial condition and the above parameters, the evolution of the system could be simulated.

At first let's consider the homogeneous case, i.e. $\beta_i=\beta=3$. From the discussion in last sector, we know that the best scale of the firm is $K$=3, and the corresponding product is $Y$=108. We have got the following simulation results.

1. Evolution results for different saving rate parameter $s_0$. Figure 1(a) shows the evolution of the market structure when $s_0$=0.2. It could be seen that in order to cover the potential demand, the firms created gradually and at last stabilized at $N$=36. The capital stock and product of each firm are: $K$=2.95, and $Y$=106.2. It almost reaches the best scale. But when $s_0$=0.05, the firms grow slower than before, and more producers have the chance to enter the market. The market structure stabilized at $N$=52 (as shown in Fig. 1(b)). The capital stock and product of each firm are only: $K$=2.21, and $Y$=73.9. The market runs in an inefficient situation. That means the final equilibrium state is not unique. It is path dependent and has the possibility of inefficiency. These are the basic properties revealed by nonlinear dynamics for economic systems.

2. Let's turn to the effects of technological progress. Here we describe the level of technology by parameter $\beta$. As we discussed in section 1, parameter $\beta$ is directly related with the scale effect of the production. The enterprises tend to have larger scale with the increasing of technological level. In the following discussion we set the saving rate coefficient $s_0=0.2$, and induce the sudden technical improvement when t=100. The results of the evolution of market structure are shown in Figure 2 and Figures 3.

Fig. 1.

When t=100, the market has almost reached its equilibrium. There have been 39 firms in the market. We let 10 firms improve their technological level suddenly, which is to change parameter $\beta$ from $\beta=3$ to $\beta=4$. As the results of technical progress and competition, supply has been enlarged and the price has gone down. But the technology progress hasn't induced the structural change of the market. All 39 firms still exist in the market but they are run in the low efficiency (shown as Fig.2 (a1) and (a2)). As shown in Fig. 2(b), in another simulation, 37 firms finally stabilized in the market. Even if 14 firms improve their technology (change parameter $\beta$ from $\beta=3$ to $\beta=4$), the market structure could not been changed, although the business cycle is more obviously as the results of competition. As shown in Figure 2 (c), when 15 firms improve their technology from $\beta=3$ to $\beta=4$, the market structure has been changed at last. All the other 22 firms with lower technology crashed as the results of competition. But because the left 15 firms can't meet the demand, there are 4 new producers with $\beta=3$ entering the market. So the market structure is formed by 19 firms including 15 with $\beta=4$ and 4 with $\beta=3$.

Fig.2

Fig.3

If when t=100, 20 producers improved their technology from $\beta=3$ to $\beta=4$ (as shown in Fig. 3), or even less producers improved their technology from $\beta=3$ to $\beta=5$, the market at last occupied by these oligarch with same share. As the results of technological progress and competition the price is lower and the demand is larger than before.

3. The competition of heterogeneous producers. In the above discussion, all or partial of producers are assumed to be identical. They have same production function with same parameters of technology. Now let's turn to the more realistic heterogeneous case that is every producer product with different technology. In our simulation, we assume that when a new producer enters the market, its scale parameter β is a real number randomly taken from 1 to 5. Other parameters and initial conditions are all the same with the above simulation.

Fig.4

From the results of numerical simulation (as shown in Fig. 4), we have found that from the time period 1 to 100 there are 32 producers has entered the market. But as the results of competition, only 23 have left. The minimum technical level for the alive firms is β=2.06. All the other firms with lower technology (β=1 to β=2.06) are failed in the competition. The average technical level for existing producers is ATL=3.30. When $t$=100, we induce a sudden technical progress, the parameter β of all the existing producers add a random number from 0 to 2. As the results of technical progress and competition, the price goes down first. And then the producers with lower technology level are get rid of from the market gradually (Here we have no sudden crash as in the homogeneous case because of the heterogeneity). At last only 9 top producers (minimum technology is β=4.35) exist in the market. The average technology level increased to β=5.27. These results give a nice description for the market competition and the industrial structure change.

One thing should be mentioned in the above simulations. Because the stochastic factors in the model, different simulations may have different final quantitative results. But the qualitative properties are unchanged.

### III. GENERALIZED MODEL WITH DIVERSITY OF DEMAND

The firm size distribution within an industry indicates the degree of industrial concentration. As mentioned in the first section, one of the stylized facts of the industrial structure is the skewed distribution of firm size. Such skewed distribution is usually described by lognormal distribution since Gibrat and its upper tail has been described by Pareto distribution or Zipf's law. Recently, empirical researches have been done to investigate the properties of firm size distribution (Stanley, Buldyrev, Falvlin. et al, 1995; Ramsden, Kiss-Haypál, 2000, Gaffeo, Gallegati, Palestrini, 2003; Fujiwaraa, Guilmib, et al, 2004). Axtell reveals that the U.S. firm size is nicely described by power law distribution (Axtell, 2001). Based on these empirical results, many theoretical analyses have been done to study the mechanism of the formation of industrial structure. Axtell has argued that complexity approach should be used to deal with the problem. Agent based modeling together with evolutionary dynamics should be helpful (Axtell, 2001). Empirical and theoretical studies on the firm size distribution have also become an important topic of econophysics (Zheng, Rodgersa, Hui, 2002; Gupta, Campanha, 2002).

In order to reproduce the skewed distribution of firm size under the framework of above model, two important factors have been introduced in the following discussion. The first is the diversity of the demand. It is related to the heterogeneity of the firms. We know that even in the same sector of the industry, the product is not exactly the same. So we assume a number of competing firms produce functionally similar products. Another point is that the decision of firms is more adaptive. We have already known that technology plays a key role in firm growth and industrial evolution. So besides investing in capital stock to expand productive capacity for existing products, the firms have chance to invest to improve their technology as well as to establish a new plant to produce new product based on the evaluation of the market.

1. Diversity of Demand

We assume that there are $N$ kinds of products in an industrial sector. At any time period $t$, each demand is denoted by $D(j,t)$. They are normal distributed in the initial. That is in the equation (8)

$$D_0(j) = D_m \exp \frac{-(j - N/2)^2}{2\sigma^2}. \tag{11}$$

In the simulation, the parameters are: $D_m$=12000, $N$=400, and σ=40. The price of $j$th product is $p_j$. It is adjusted with the same mechanism as in Section II. But it is only determined by $j$th supply and demand. The initial price for each product is given by $p(j,0) = 0.12 - 0.04 j/N$. In equation (9) the parameters $p_{0i}$ that determine the adjustment of the price are $p_0(j) = 0.15 - 0.04 j/N$.

For $i$th firm, at time period $\tau$, it has established a plant to produce product $j$. Its capital stock and product are denoted by $K(i,j,\tau,t)$ and $Y(i,j,\tau,t)$. And the firm size is described by the total production or total capital stock belongs to the same firm. They are $Y(i,t) = \sum_{j,\tau} Y(i,j,\tau,t)$ and $K(i,t) = \sum_{j,\tau} K(i,j,\tau,t)$. The aggregate demand and supply are given by: $D(t) = \sum_{j} D(j,t)$ and $Y(t) = \sum_{i} Y(i,t)$.

2. Adaptive decision making of the firms

At the beginning of each time period, every plant will decide its saving rate according to his knowledge about the situation of the market:

$$s(j,t) = s_0 (1.0 - \exp(5 \frac{Y(j,t) - D(j,t)}{D(j,t)})) . \tag{12}$$

Then he will make his investment decision based on the evaluation of expected returns. Let the new investment be

$$\Delta K(i,j,\tau,t) = s(j,t-1) R(i,j,\tau,t-1) - \delta K(i,j,\tau,t-1) . \tag{13}$$

It could be used for:

A. Increasing capital stock to enlarge the productivity capacity.

B. Investing in R&D sector to improve his technology. In our model, the technology is described by parameter $\beta$ and it relate to the scale effect directly. We assume that the increase of technology for a given investment follows

$$\begin{aligned} \beta(i,j,\tau,t) = &\beta(i,j,\tau,t-1) \\ &+ \Delta K(i,j,\tau,t)\bigl(\beta(i,j,\tau,t-1) - 2.9\bigr)\bigl(1 - \beta(i,j,\tau,t-1)/5\bigr) \end{aligned} \tag{14}$$

It is a sigmoid curve with limit $\beta_{max}=5.0$.

C. If the new investment is greater than 1, the firm can also determine to establish a new plant in the $j$th product with maximum difference between demand and supply. The technical parameter of the new plant is the same as the investor.

Fig.5

Fig.6

3. Born and death of the firms.

If the total supply can not match the total demand, some new producers will enter the market. A new firm with production *j* will enter the market with probability

$$q(j,t) = q(D(j,t-1) - Y(j,t-1))/D(j,t-1), \text{ when } D(j,t\text{-}1) > Y(j,t\text{-}1). \tag{15}$$

Where *q* is a parameter in [0,1].

Then we have done a lot of numerical simulations under different sets of parameters. Figures 5 and 6 give two of the results including the evolutionary dynamics of aggregate supply and demand, average price, the number of firms, and the final steady distribution of firm size. The firm size is described by the total production or total capital stock of the firms. All the graphs for the distribution are the results of 10 simulations under the same parameters set. The size distribution of firms is normally a skewed distribution that can be described by power law. In the log-log Zipf plot, the distribution curves are all near a straight line. From the simulation, we have found that the final distribution is robust to a certain extent under the reasonable parameter regions. Figure 7 shows the several situations. The following factors take an important role in the final skewed distribution of firm size: 1. Self-reinforcing mechanism caused by increasing returns, that may results in first-mover advantages and path-dependent; 2. Firm strategies change adaptively as the firm grows and competition evolves, including investment in new products and technical progress; 3. Demand diversities or widespread heterogeneity in the technological capabilities of different firms.

It is interesting to know the strategies adopted by a firm during the market evolution. The results are shown in Figure 8.

Fig. 7

Fig.8

## V. CONCLUDING REMARKS

The formation of industrial structure is a very interesting topic for evolutionary economic research. It is also an attracting area for the study of pattern formation. The model presented here using the multi-agent approach gives us a dynamic perspective for the evolution of industrial structure. It derives some generic properties of the underlying competition process. The corresponding results can help us reaching a deep understanding for the related problems, such as the effects of technical progress and the global behavior with the change of structure.

There are several problems remain for further research. 1. The mechanism for the adjustment of price and demand should be more realistic. Other functions (linear or nonlinear) for the relationship of price and demand should be adapted. So we can discuss rigid or elastic price adjustment and the effects of the price elasticity of the demand. 2. The production function should also include the change of labor. So we can take the change of employment into account. 3. The contestable market could be also a more realistic one. We should consider a market in growth. And the entry or not of producer is determined by the expected returns instead of by the shortage of supply. 4. Other stylized facts of industrial organization should be investigated, especially the growth rate of the firm.

**Figure Captions**

Fig.1 Evolution of market structure under different saving rate $s_0$. The changing of demand (D), supply (S), price level (P), and number of firms (N) are labeled near the corresponding curves. (a) $s_0$=0.2. (b) $s_0$=0.05. (c) Growth of capital for a firm in the above two cases.

Fig.2 Evolution of the market after the technical progress. There is threshold effect in the system. The demand (D), supply (S), price level (P), and number of firms (N) are labeled near the corresponding curves. (a) 10 firms change $\beta$ from $\beta$=3 to $\beta$=4, (a1) shows the change of aggregate economic variables including the number of firms, (a2) shows the change of capital of firm No.5 and No. 25; (b) 14 firms change $\beta$ from $\beta$=3 to $\beta$=4; (c) 15 firms change $\beta$ from $\beta$=3 to $\beta$=4.

Fig.3 Evolution of the market after the technical progress. 20 producers improve their technology from $\beta$=3 to $\beta$=4. (a) The change of aggregate economic variables including the number of firms. (b) The change of capital of firm No.5 and No. 25. Firm No. 25 was failed in the competition of the market.

Fig.4 Evolution of heterogeneous producer. (a) Price (P), demand (D) and supply (S), (b) Number of firms (N) and average level of technology (ATL).

Fig.5 Evolution of the aggregate demand (D), supply (S), average price level (P), and number of firms (a). And (b) shows the log-log Zipf plots of the final size distribution of firms described by total production (b1) and total capital stock (b2). Probability for the birth of new firm is $q$=0.01, and $s_0$=0.2.

Fig.6 Evolution of the aggregate demand (D), supply (S), average price level (P), and number of firms (N) (a), and the final size distribution of the firms described by total capital stock (b). Probability for the birth of new firm is $q$=0.1, and $s_0$=0.25. The total number of firms is around 420

Fig.7. Log-Log Zipf plots for the final size distribution of the firms described by total production under different set of parameters. (a) Probability for the birth of new firm is $q$=0.01, and $s_0$=0.3. The total number of firms is around 200. (b) Probability for the birth of new firm is $q$=0.6, and $s_0$=0.25. The total number of firms is around 670. (c) Probability for the birth of new firm is $q$=1.0, and $s_0$=0.2. The total number of firms is around 720.

Fig.8. Strategies evolution of a firm.

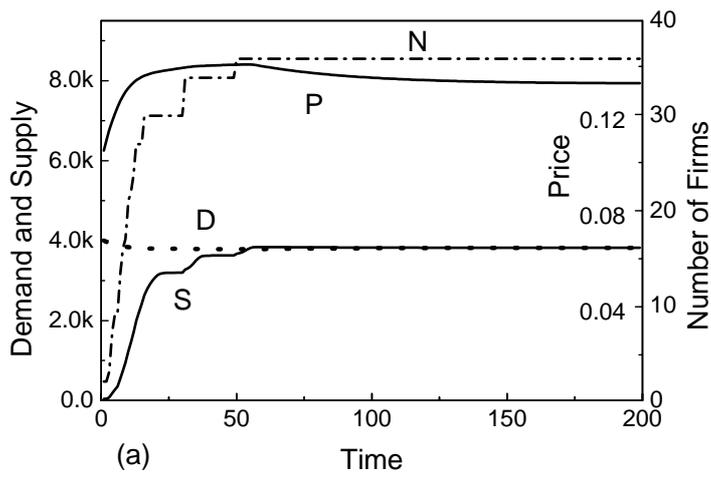

(a)

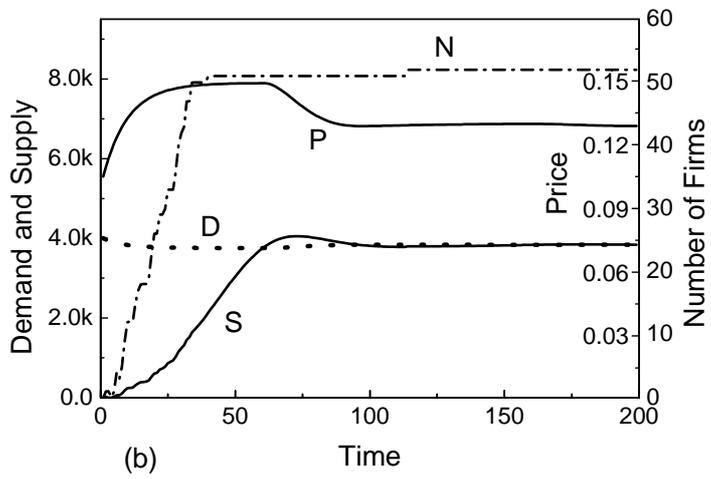

(b)

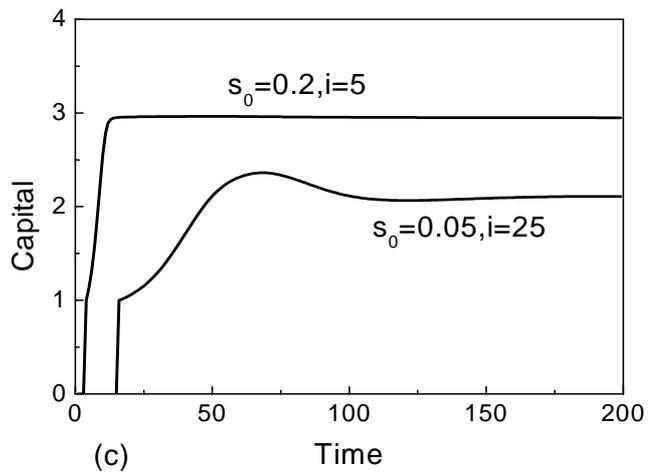

(c)

**Fig. 1**

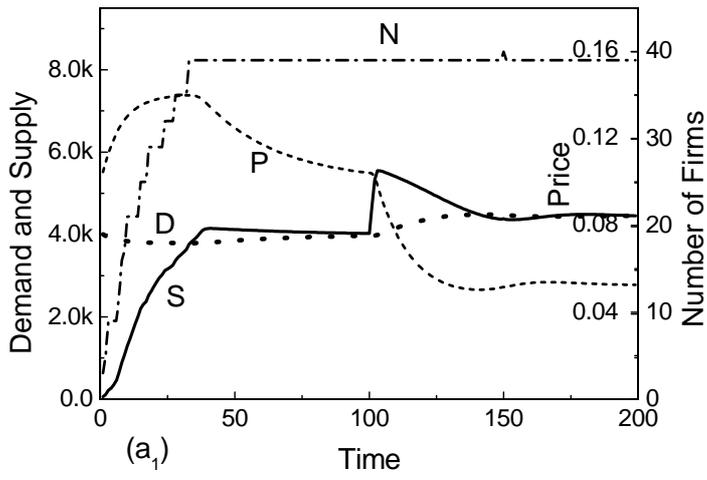

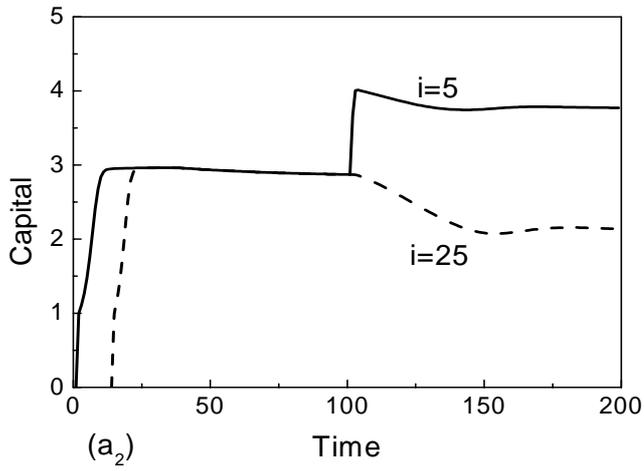

**Fig. 2 (a1) (a2)**

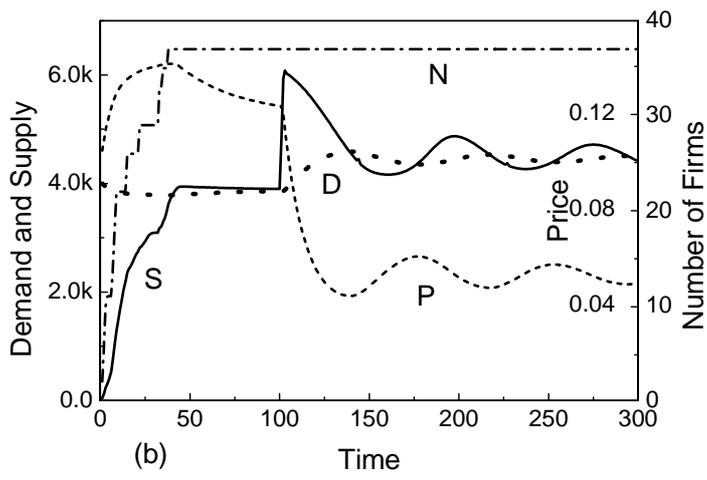

(b)

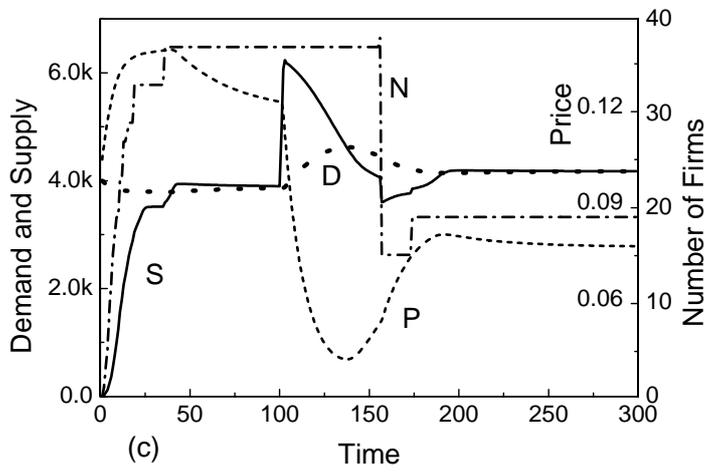

(c)

**Fig. 2 (b), (c)**

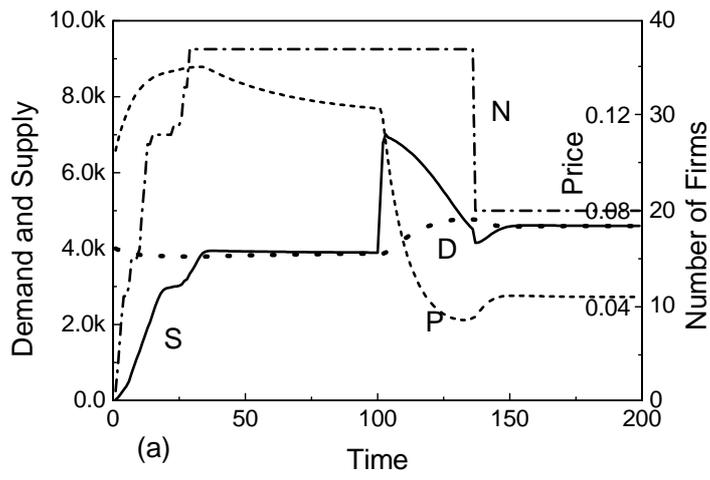

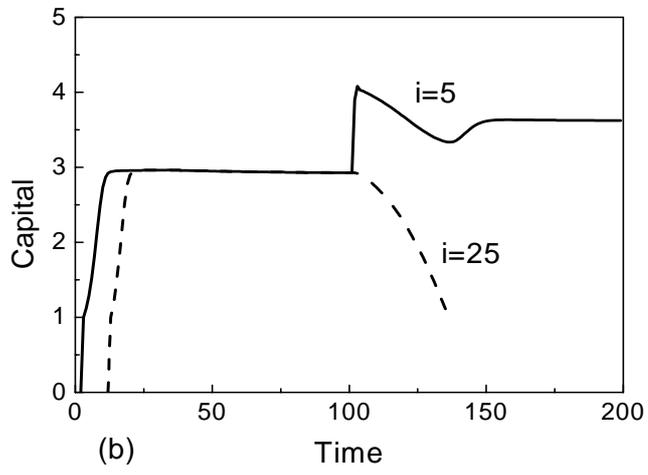

**Fig. 3**

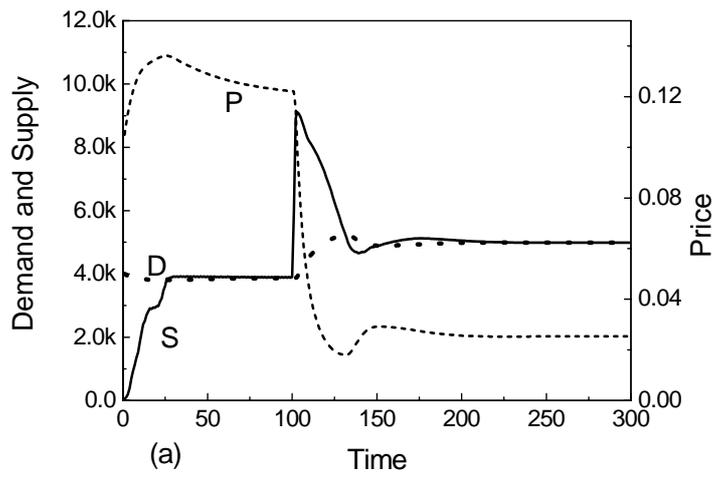

(a)

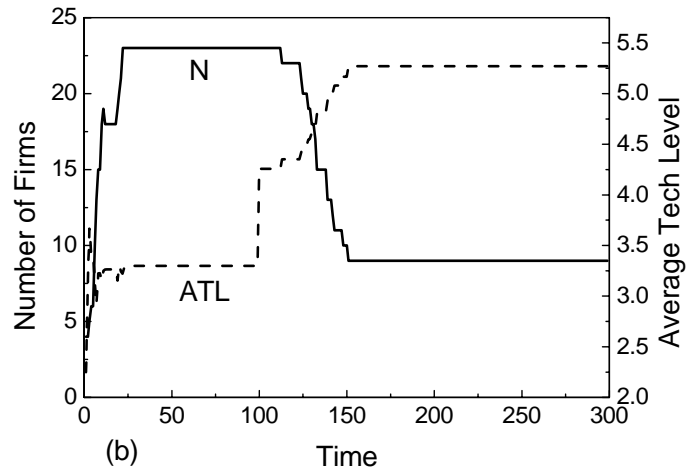

(b)

**Fig. 4**

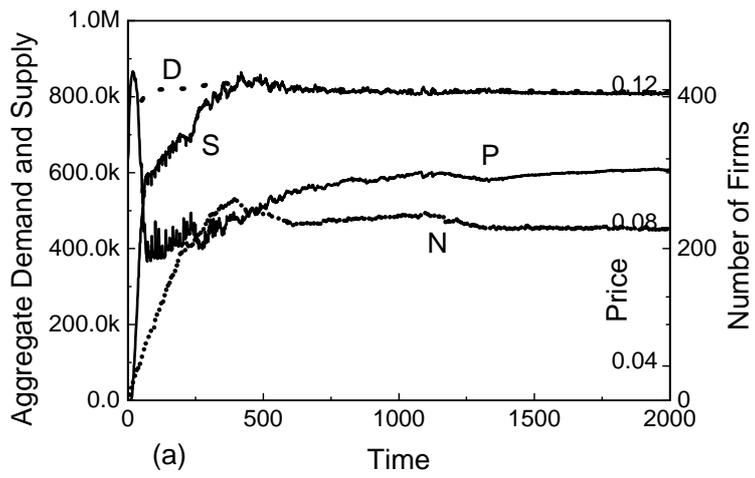

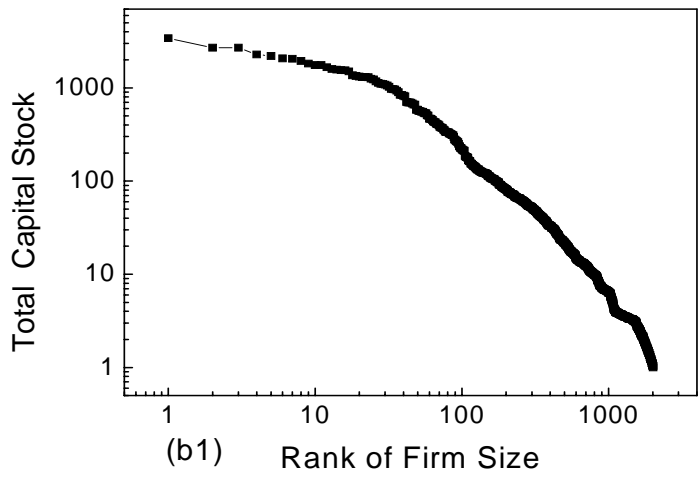

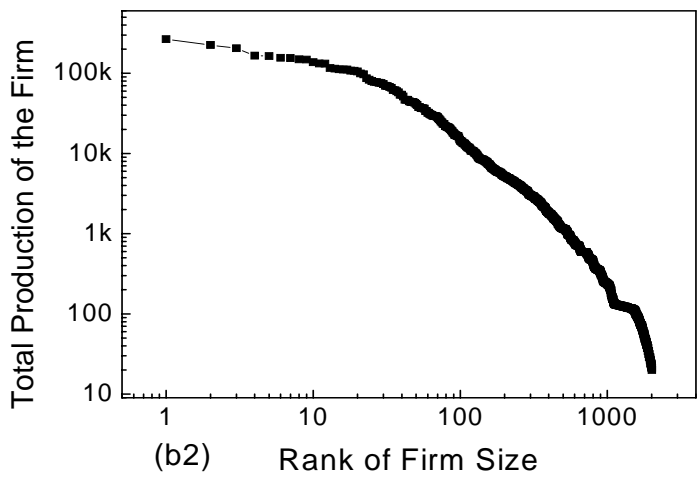

**Fig. 5**

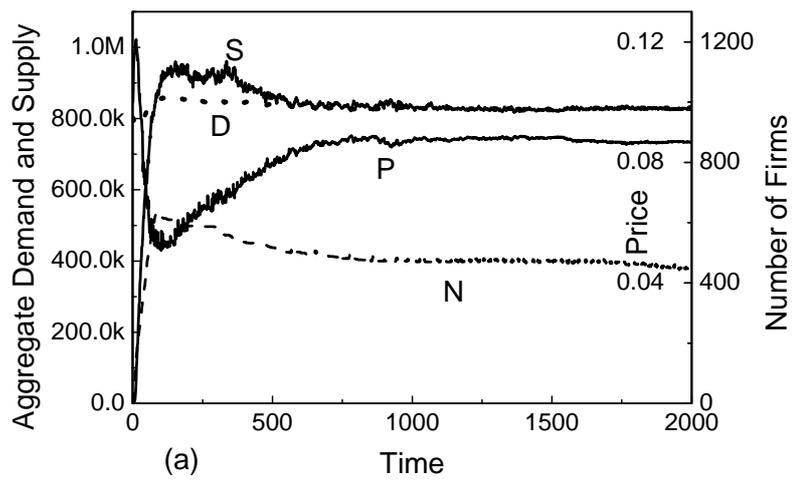

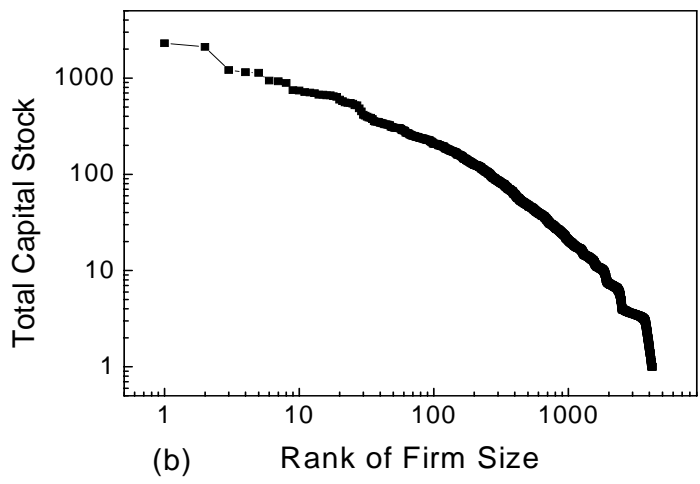

**Fig.6**

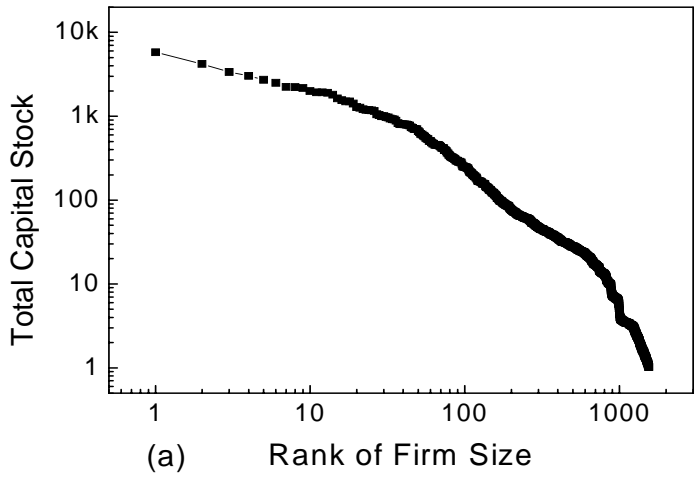

(a) Rank of Firm Size

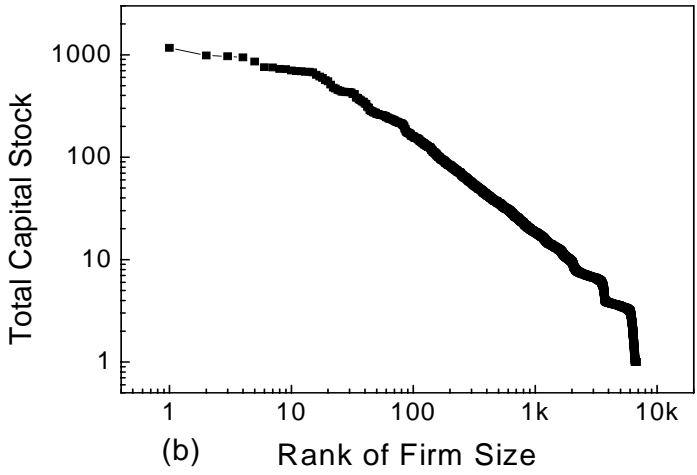

(b) Rank of Firm Size

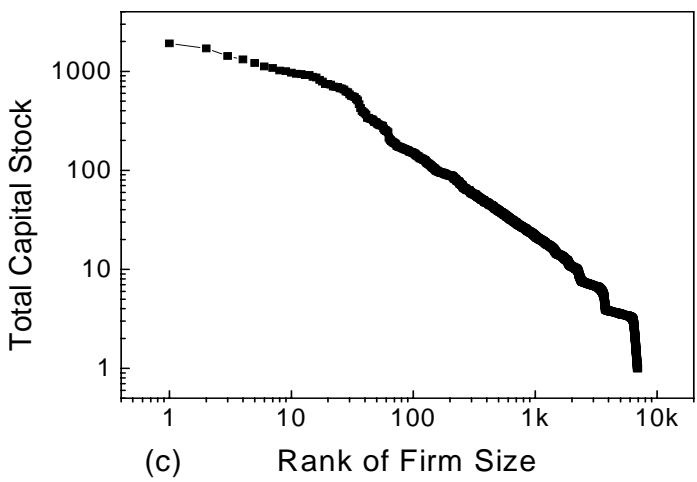

(c) Rank of Firm Size

**Fig.7**

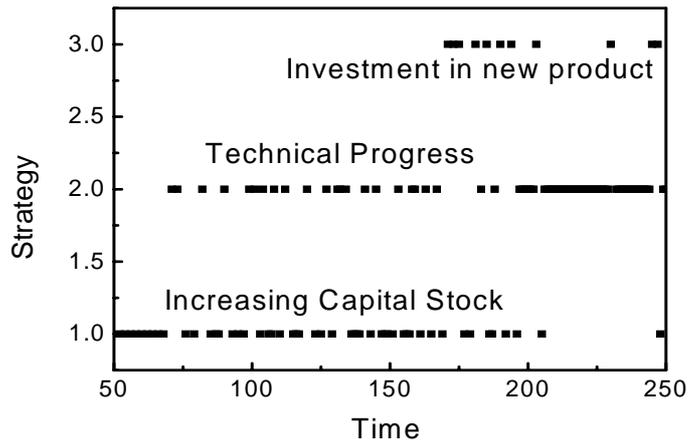

**Fig. 8**